\documentclass[a4paper,english]{lipics-v2019}
\usepackage{nth}
\usepackage{microtype}
\usepackage{breqn}
\usepackage{ifthen}
\usepackage{placeins}
\hideLIPIcs
\nolinenumbers

\newcommand{\old}[1]{{}}

\title{Computing Convex Partitions for Point Sets in the Plane: The CG:SHOP Challenge 2020}
\titlerunning{Convex Partitioning: CG Challenge 2020}

\author{Erik D.~Demaine}{CSAIL, MIT, USA}{edemaine@mit.edu}{}{}
\author{Sándor P.~Fekete}{Department of Computer Science, TU Braunschweig, Germany}{s.fekete@tu-bs.de}{https://orcid.org/0000-0002-9062-4241}{}
\author{Phillip Keldenich}{Department of Computer Science, TU Braunschweig, Germany}{p.keldenich@tu-bs.de}{https://orcid.org/0000-0002-6677-5090}{}
\author{Dominik Krupke}{Department of Computer Science, TU Braunschweig, Germany}{d.krupke@tu-bs.de}{https://orcid.org/0000-0003-1573-3496}{}
\author{Joseph S.~B.~Mitchell}{Department of Applied Mathematics and Statistics, Stony Brook University, USA}{joseph.mitchell@stonybrook.edu}{https://orcid.org/0000-0002-0152-2279}{}
\authorrunning{E.~D.~Demaine, S.~P.~Fekete, P.~Keldenich, D.~Krupke, J.~S.~B.~Mitchell}
\Copyright{E.~D.~Demaine, S.~P.~Fekete, P.~Keldenich, D.~Krupke, J.~S.~B.~Mitchell}

\ccsdesc{Theory of computation $\rightarrow$ Computational geometry}
\ccsdesc{Theory of computation $\rightarrow$ Design and analysis of algorithms}

\keywords{Computational Geometry, geometric optimization, complexity, convex partition, algorithm engineering, contest}

\supplement{}

\funding{} 

\acknowledgements{}

\EventAcronym{CG:SHOP 2020}

\begin{document}
\maketitle
\begin{abstract}
We give an overview of the 2020 Computational Geometry Challenge,
which targeted the problem of partitioning the convex hull of a given planar point
set $P$ into the smallest number of convex faces, such that no point of $P$
is contained in the interior of a face. 
\end{abstract}

  \section{Introduction: Solving Hard Optimization Problems}
Computational Geometry is the study of algorithms for solving
problems on geometric data. Geometric problem arise in numerous applications
from fields that include Robotics (motion planning and
visibility problems), Operations Research (geometric location and search, route
planning), Integrated Circuit Design (IC geometric design and verification),
Computer-Aided Engineering (CAE), Mesh Generation, Computer Vision and Shape
Reconstruction. These problems are often cast as optimization problems in which
one is required to find a set of specific geometric objects that maximizes or
minimizes a given objective function.  

In the roughly 40 years since the beginning
of Computational Geometry, a wide range of optimization problems have been
proposed and investigated by computational geometers, mainly from a theoretical
point of view. Many of those tasks belong to the NP-hard class of problems, for
which the existence of polynomial-time algorithms implies P=NP. While
Computational Geometry has considered a wide spectrum of NP-hard optimization
problems, positive results typically imply polynomial-time, constant-factor
approximation algorithms, without much regard for practical solution quality,
realistic running times, or even exact solutions.

Even before the area of Computational Geometry was started, researchers from
the mathematical field of Combinatorial Optimization were already very
successful in tackling large instances of hard problems, in particular, based
on Integer Linear Programming. However, work from this field has mostly focused
on discrete structures, such as graphs. This differs from geometric problems,
which typically require a deeper understanding and modeling of a wide range of
continuous structures: geometric structures are not automatically discrete, and
additional geometric computations may be prohibitively expensive. Moreover,
even aspects of geometric optimization problems that are theoretically ``easy'',
because they allow a polynomial-time solution, may be problematic in practice,
because they need to be treated again and again during the process of solving a
difficult optimization problem. This makes it important to tune algorithmic
solutions, and often combine them with appropriate geometric data structures
for efficient practical computation. In principle, this is the approach taken
by the comparatively newer area of Algorithm Engineering; however, at this point the
impact on hard problems treated in the community of Computational Geometry has
been somewhat limited, as the focus has been more on streamlining computational
efficiency, rather than exact methods. Combining approaches from all these
different communities is highly desirable.

Scientific progress is largely triggered by considering new problems. A
particularly attractive way of motivating a wide range of researchers to work
on new challenges is to pose interesting competitions: These ensure that
results are achieved in a timely fashion, the practical state of the art gets
visibly established, and credit for success is given in a visible manner.  On
the theoretical side of computational geometry, these objectives have lead to
The Open Problems Project (TOPP), maintained by Demaine, Mitchell and O’Rourke~\cite{TOPP},
a library of unsolved problems, rather than instances. On the more
practical side of combinatorial optimization, there have been different angles
to motivate practical efforts. Each benchmark library (such as the TSPLIB~\cite{TSPLIB}) 
by itself constitutes a collection of challenges. Since 1990, the DIMACS
implementation challenges have addressed questions of determining realistic
algorithm performance where worst-case analysis is overly pessimistic and
probabilistic models are too unrealistic. Since 1994, the Graph Drawing (GD)
community has held annual contests in conjunction with its annual symposium to
monitor and challenge the current state of the graph-drawing technology and to
stimulate new research directions for graph layout algorithm. 

\old{
At the Computational Geometry Week 2019 (which
includes SoCG, the Annual Symposium on Computational Geometry), we ran a
``CG:SHOP Challenge'' (Computational Geometry: Solving Hard Optimization Problems), 
based on the problem of computing
minimum-area polygons.  This Challenge was well received within the community
and also brought in a number of teams from other areas that are more concerned
with optimization. As one outcome, the community decided to give this CG
Challenge space both in the CG Week program and the SoCG proceedings. In
addition, we chose to expand the scope and timeline of the event to allow using
it as the basis for student projects and practical work by junior researchers.

This survey presents content and outcomes of this second CG Challenge.
}

Until recently, there have not been any comparable challenges in the context of
Computational Geometry.
%
%
The ``CG:SHOP Challenge'' (Computational Geometry: Solving Hard
Optimization Problems) originated as a workshop at the 2019
Computational Geometry Week (CG Week) in Portland, Oregon in June,
2019.  The goal was to conduct a computational challenge competition
that focused attention on a specific hard geometric optimization
problem, encouraging researchers to devise and implement solution
methods that could be compared scientifically based on how well they
performed on a database of instances. While much of computational
geometry research has targeted theoretical research, often seeking
provable approximation algorithms for NP-hard optimization problems,
the goal of the CG Challenge was to set the metric of success based on
computational results on a specific set of benchmark geometric
instances. The 2019 CG Challenge focused on the problem of computing
minimum-area polygons whose vertices were a given set of points in the
plane.  This Challenge generated a strong response from many research
groups, from both the computational geometry and the combinatorial
optimization communities, and resulted in a lively exchange of
solution ideas.

For CG Week 2020, the second CG:SHOP Challenge became an event within
the CG Week program, with top performing solutions reported in the
Symposium on Computational Geometry proceedings. The schedule for the
Challenge was advanced earlier, to give an opportunity for more
participation, particularly among students, e.g., as part of course
projects.

\old{ Joe's version:
The ``CG:SHOP Challenge'' (Computational Geometry: Solving Hard
Optimization Problems) originated as a workshop at the 2019
Computational Geometry Week (CG Week) in Portland, Oregon in June,
2019.  The goal was to conduct a computational challenge competition
that focussed attention on a specific hard geometric optimization
problem, encouraging researchers to devise and implement solution
methods that could be compared scientifically based on how well they
performed on a database of instances. While much of computational
geometry research has targeted theoretical research, often seeking
provable approximation algorithms for NP-hard optimization problems,
the goal of the CG Challenge was to set the metric of success based on
computational results on a specific set of benchmark geometric
instances. The 2019 CG Challenge focussed on the problem of computing
minimum-area polygons whose vertices were a given set of points in the
plane.  This Challenge generated a strong response from many research
groups, from both the computational geometry and the combinatorial
optimization communities, and resulted in a lively exchange of
solution ideas.

For CG Week 2020, the second CG:SHOP Challenge became an event within
the CG Week program, with top performing solutions reported in the
Symposium on Computational Geometry proceedings. The schedule for the
Challenge was advanced earlier, to give an opportunity for more
participation, particularly among students, e.g., as part of course
projects.
}

\section{The Challenge: Minimum Convex Partitions for Planar Point Sets}
\subsection{The Problem}
The specific problem that formed the basis of the 2020 CG Challenge was the following:

\textbf{Problem} {\sc Minimum Convex Partition} (MCP) in the plane.

\textbf{Given:} A set $P$ of $n$ points in the plane. 

\textbf{Goal:} A plane graph with vertex set $P$ (with each point in $P$ having positive degree) that
partitions the convex hull of $P$ into the smallest possible number $u(P)$ of convex
faces. 

\medskip
Note that collinear points are allowed on face boundaries. Each internal face angle at each point of $P$ is at most $\pi$.

\subsection{Related work}
The problem of computing a partition of a simple polygon, having $n$ vertices, into a minimum number of convex pieces has been well studied. If Steiner points are allowed to be added, then Chazelle and Dobkin~\cite{chazelle1979decomposing} gave an algorithm, based on dyanamic programming, which computes exactly an optimal solution in time $O(n+r^3)$, where $r$ is the number of reflex vertices of the given polygon.
For decompositions using only diagonals (no Steiner points),
Greene~\cite{greene1983decomposition} gave an $O(r^2n^2)$ algorithm, also based
on dynamic programming.
Keil~\cite{keil1985decomposing} improved the running time to $O(rn^2\log n)$ and gave a proof of NP-hardness
for polygons with holes; later, Keil and Snoeyink~\cite{keil2002time} improved the complexity
to $O(n+r^2\min\{r^2,n\})$.

The complexity of {\sc Minimum Convex Partition} (MCP) in the
plane was unknown
when the 2020 CG Challenge began in September 2019.  In November 2019, 
Grelier announced a proof of NP-hardness for the case of
planar point sets in not necessarily general position~\cite{grelier2019minimum}.
The complexity of the MCP for points in general position is still open at the time of this writing.
On the positive side, a number of positive algorithmic results have been known for a while,
assuming special properties of $P$. 
For point sets that can be decomposed into a limited number of
convex layers, Fevens, Meijer and Rappaport~\cite{fevens2001minimum} gave a
polynomial-time algorithm. Assuming that no three points
are collinear, Knauer and Spillner~\cite{knauer2006approximation} gave a 3-approximation algorithm
that runs in $O(n\log n)$, and a $\frac{30}{11}$-approximation of complexity $O(n^2)$.

Worst-case bounds have also been considered; for this purpose, define
$U(n)$ as the maximum number $u(P)$ over all non-degenerate point sets $P$ with $n\geq 3$.
In 1998, Urrutia~\cite{urrutia1998open} conjectured that $U(n)\leq n+1$.
Neumann-Lara, Rivera-Campo and Urrutia~\cite{neumann2004note} showed that
$U(n) \leq \frac{10n-18}{7}$; Lomeli-Haro~\cite{lomeliharo2012minimal} showed that $U(n) \leq \frac{10}{7}n - h$, where $h$ is the number of points on the convex hull. This bound was improved by Hosono~\cite{hosono2009convex}  
to $U(n)\leq\frac{(7(n-3)}{5}$, and later by Sakai and Urrutia~\cite{sakai2019convex} to 
$U(n)\leq\frac{4}{3}-2$. Conversely, Knauer and Spillner~\cite{knauer2006approximation}
showed that $U(n)\geq n+2$, which was improved by Garc\'ia-L\'opez and Nicol\'as~\cite{garcia2013planar}
to $U(n)\geq \frac{12}{11}n-2$.

Aiming for solutions to benchmark instances, Barboza, Souza and Rezende~\cite{barboza}
gave an integer linear programming formulation of the MCP, and showed that
this can be used to solve instances with up to 50 points to provable optimality.

\subsection{Instances}
The contest started with a total of 247 benchmark instances, as follows.
Each of these instance consisted of $n$ points in the plane with integer coordinates.
For $n\in\{10$, $15$, $20$, $25$, $30$, $35$, $40$, $45$, $50$, $60$, $70$, 
$80$, $90$, $100$, $200$, $300$, $400$, $500$, $600$, $700$, $800$, $900$, $1000$, $2000$, $3000$, $4000$, $5000$, $6000$, $7000$, $8000$,
$9000, 10000, 20000, 30000, 40000, 50000, 60000, 70000, 80000$, $90000, 100000\}$,
there were six instances each.  In addition, there is one instance of size 
$n=1000000$.

The instances were of four different types:

\begin{itemize}
\item
\textbf{uniform:} uniformly at random from a square
\item
\textbf{edge:} randomly generated according to the distribution of the rate of change (the ``edges'') of an image
\item
\textbf{illumination:} randomly generated according to the distribution of brightness of an image (such as an illumination map)
\item
\textbf{orthogonally collinear points:} randomly generated on an integral grid to have a lot of collinear points (similar to PCBs and distorted blueprints).
.
\end{itemize}

These instances were based on point sets that were originally
generated for the 2019 Challenge; as such, they tended to be in general
position. To account for this and the progress in complexity (which was based on
instances with collinear points), a further 99 instances with larger numbers 
of collinear points were added in January 2020.

\subsection{Evaluation}
The comparison between different teams was based on an overall \emph{score}.
For each instance, this score is a number between 0 and 1, with higher 
scores corresponding to better solutions. The trivial solution, i.e., 
a triangulation, corresponds to a score of 0, and a solution 
without any edges, which is, of course, infeasible, corresponds to a score of 1.

For an instance, i.e., a point set $P$ consisting of $n$ points, let $c$ be the
number of points on the convex hull of $P$. Observe that any triangulation of
$P$ is a convex partition with $2n-2-c$ bounded faces and $3(n-1)-c$
edges. Moreover, any convex partition $\Pi$ can be obtained by starting with a
triangulation containing its edges, and removing the excess edges one by one.
In this process, removing a single edge also decreases the number of bounded
faces by exactly 1. Thus, any solution $\Pi$ with $f = 2n-2-c-s$ faces
for some $s\geq 0$ has $m = 3(n-1)-c-s$ edges and vice versa. This
allows using the number $m(\Pi)$ of edges in a solution $\Pi$ instead of the
number of faces to determine the score of $\Pi$ as
\[\mbox{score}(\Pi) := \frac{s(\Pi)}{3(n-1)-c},\]
where $s(\Pi) = 3(n-1)-c-m (\Pi).$
In other words, the score for a solution to an instance 
$P$ is the fraction of edges removed from a triangulation of 
$P$.

The total score achieved by each team was the sum of all individual instance scores; 
only the best feasible solution submitted was used to compute the score. Participation
required submitting feasible solutions. Feasibility was checked at the time of
upload. Failing to submit a feasible solution for an instance resulted in a
default score of 0 for that instance.

In case of ties, the tiebreaker was set to be the time a specific score was obtained. 
This turned out not to be necessary.

\subsection{Categories}
The contest was run in an \emph{Open Class}, in which participants could use any
computing device, any amount of computing time (within the duration of the
contest) and any team composition. In the \emph{Junior Class}, a
 team was required to consist exclusively of participants who were eligible
according to the rules of CG:YRF (the \emph{Young Researchers Forum} of CG Week), 
defined as not having defended a formal doctorate before 2018.

The demand for an additional \emph{Limited Class}
(to be run on a specific server that was to be uniform for all participants)
turned out to be too low to justify the additional effort.

\subsection{Server and Timeline}
The contest itself was run through a dedicated server at TU Braunschweig,
hosted at \url{https://cgshop.ibr.cs.tu-bs.de/competition/cg-shop-2020/}.
It opened at  18:00 CEDT (noon, EDT) on September 30, 2019, and closed
at 24:00 (midnight, AoE), February 14, 2020. 

\section{Outcomes}
A total of 21 teams participated in the contest. In the end, the top 10 in the leaderboard
looked as shown in Table~\ref{tab:top10}; note that according to the scoring function,
a higher score is better.

\begin{table}[h!]
  \begin{center}
    \caption{The top of the final leaderboard. ``Best solutions'' are the best found by any participating team,
which does not exclude the possibility of better solutions. ``Unique best'' solutions are those that were not found
by any other team.}
    \label{tab:top10}
    \begin{tabular}{|r|l|c|c|c|} 
      \hline
      \textbf{Position} & \textbf{Team} & \textbf{Score} & \textbf{\# best }  & \textbf{\# unique best}\\
      			& 		& 		 & \textbf{solutions} & \textbf{solutions}\\
      \hline
      1 & Team UBC 	& 175.172880	& 209 & 11\\
      2 & OMEGA 	& 175.130597	& 297 & 126\\
      3 & CGA-Sbg	& 175.040207	& 187 & 0 \\
      4 & Les Shadoks	& 174.695586	& 160 & 6 \\
      5 & G-SCOP	& 174.543068	& 138 & 0 \\
      6 & Min2Win@Zurich& 174.384784	& 121 & 0 \\
      7 & TUFUnky4you	& 173.973716	& 100 & 0 \\
      8 & Team Technion & 173.621857	& 99 & 0 \\
      9 & Sapucaia, de Rezende, de Souza & 170.939574 & 88 & 0 \\
      10 & ucsbtheorylab & 169.975180	& 76 & 0 \\
      \hline
    \end{tabular}
  \end{center}
\end{table}

\begin{figure}[]
        \begin{center}
        \resizebox{.8\linewidth}{!}{\includegraphics{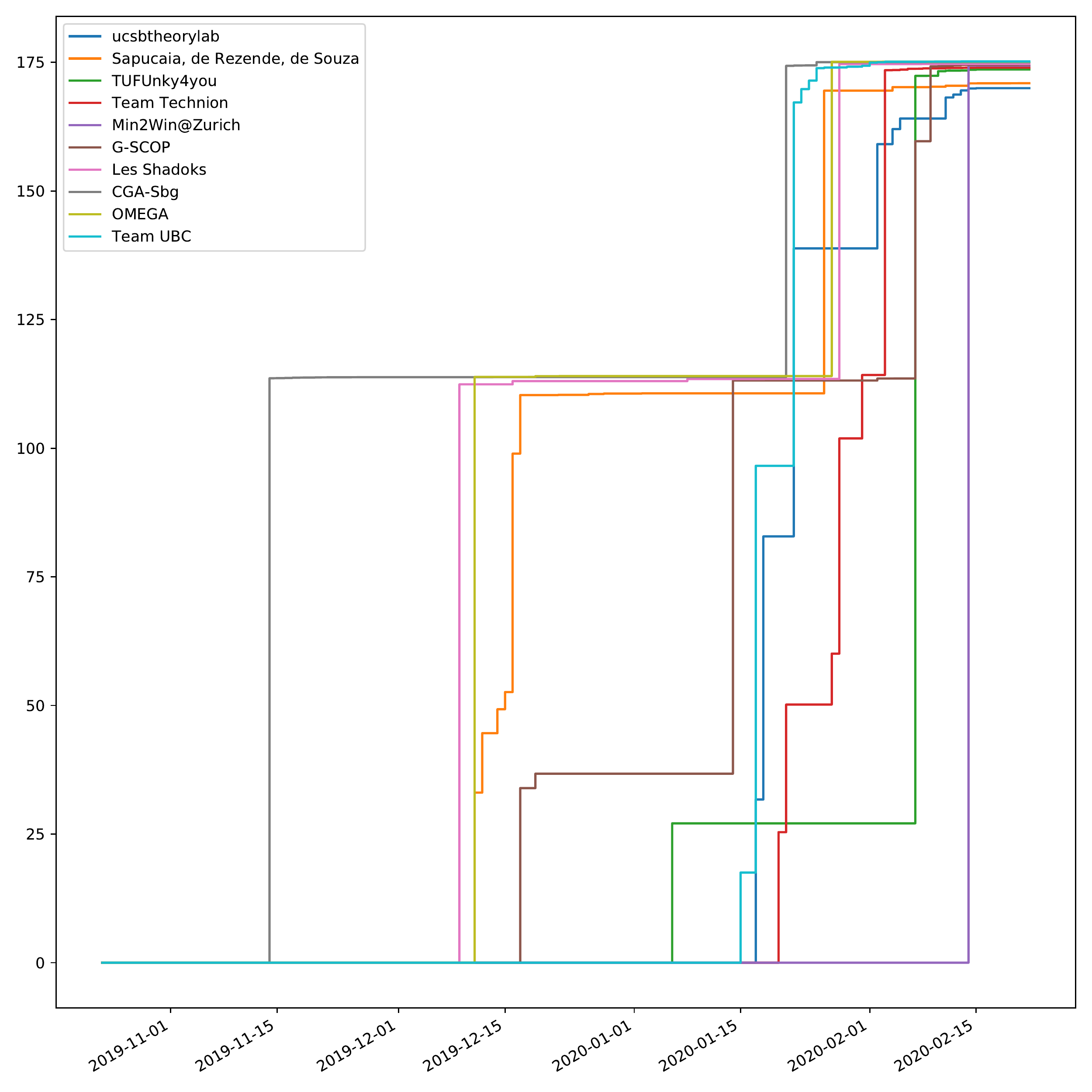}}
        \end{center}
                \caption{Total score over time for the best ten teams.}
        \label{fig:top10}
\end{figure}

\begin{figure}[]
        \begin{center}
        \resizebox{.9\linewidth}{!}{\includegraphics{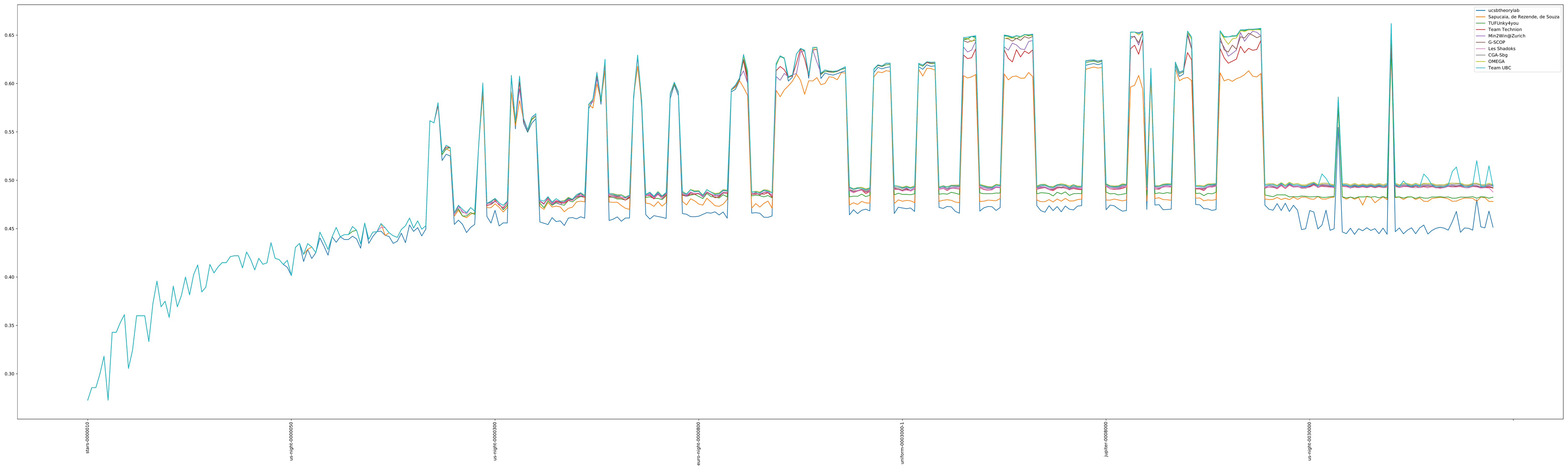}}
        \end{center}
                \caption{All instance scores for the best ten teams.}
        \label{fig:allscores}
\end{figure}

The progress over time of each team's score can be seen in Figure~\ref{fig:top10};
the best solutions for all instances (displayed by score) can be seen in Figure~\ref{fig:allscores}.

\begin{figure}[]
        \hspace*{-.15\linewidth}\resizebox{1.25\linewidth}{!}{\includegraphics{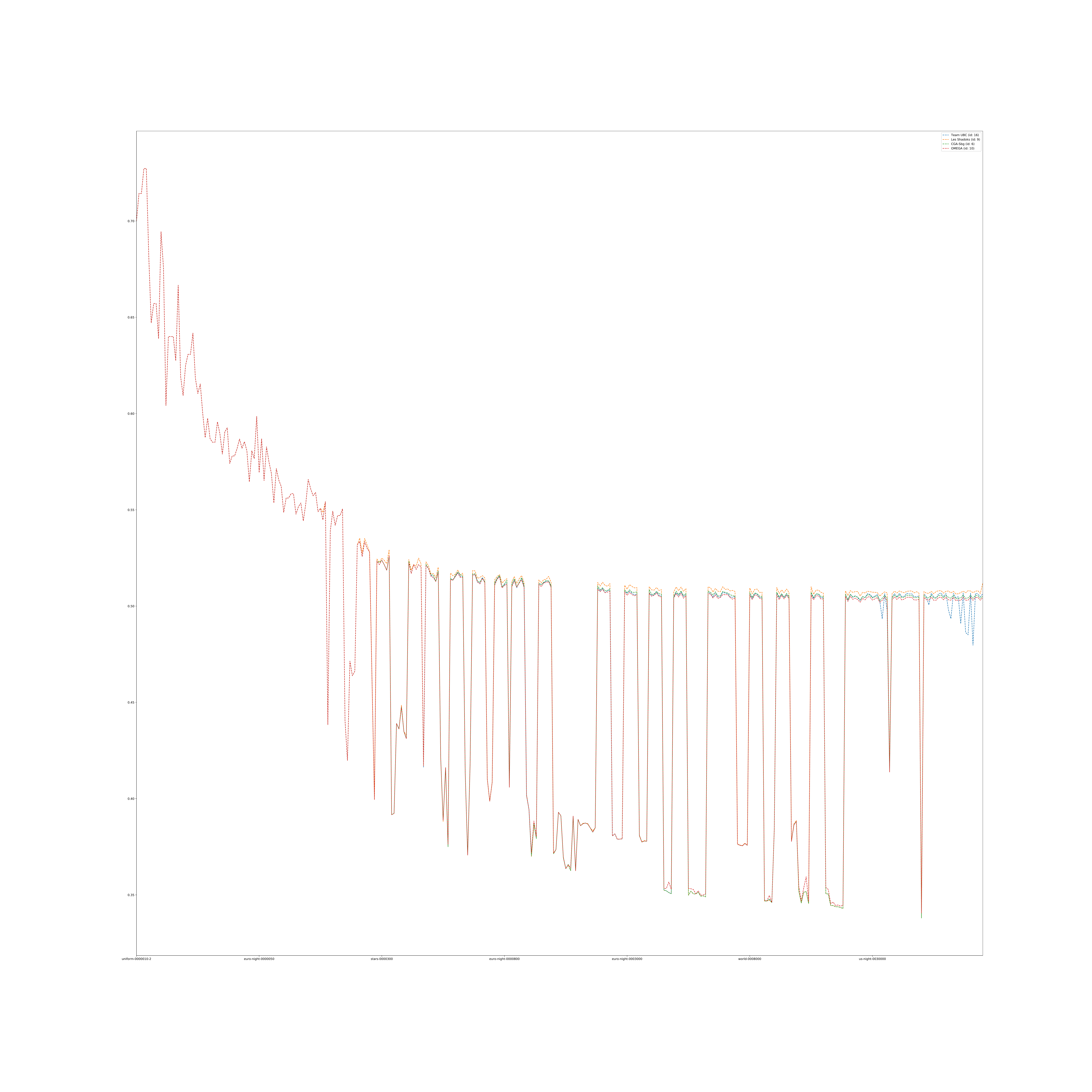}}
	\vspace*{-2cm}
                \caption{Normalized number of faces for the top 4 teams. Note how OMEGA (shown in red) achieved very slightly better
			solutions for many instances, while Team UBC (shown in blue) found significantly better solutions for a relatively
			small number of instances, which made the difference in the overall outcome.} 
        \label{fig:score}
\end{figure}

Clearly, the outcome was quite tight between the top teams; in particular, Team UBC
and Omega were only separated by 0.02\% in their respective scores. As can be seen from 
Figure~\ref{fig:score} (which shows the normalized number of faces instead of the respective
scores for the top 4 teams), OMEGA found very slightly better solutions for a large number
of instances (which is also reflected by the high number of unique best solutions in Table~\ref{tab:top10}),
while Team UBC found significantly better solution for six of the instances. The latter
was sufficient for the overall win.


The top 3 finishers in the Open Class were invited for contributions in the 
2020 SoCG proceedings, as follows.

\begin{enumerate}
\item Team UBC: Da Wei Zheng, Jack Spalding-Jamieson and Brandon Zhang~\cite{UBC}.
\item Team OMEGA: Laurent Moalic, Dominique Schmitt, Julien Lepagnot and Julien Kritter~\cite{Mulhouse}.
\item Team CGA-Sbg: G\"unther Eder, Martin Held, Stefan de Lorenzo and Peter Palfrader~\cite{Salzburg}.
\end{enumerate}

Consisting only of students, Team UBC was also the runaway winner of the Junior Class.

All three teams engineered their solutions based, broadly, on variants of local search methods, with the use of randomization and constraint programming~\cite{UBC}, genetic approaches~\cite{Mulhouse}, and tailored initial decompositions~\cite{Salzburg}.
Details of their methods and the engineering decisions they made are given in their respective papers.

\section{Conclusions}
The 2020 CG:SHOP Challenge motivated a considerable number of teams to engage in intensive optimization studies.
Not only did this lead to practical developments, it also triggered theoretical progress, as demonstrated
by Grelier~\cite{grelier2019minimum}. We are confident that this will motivate further work
on the problem of Minimum Convex Partitions, as well as other practical geometric optimization work.

\bibliography{references}
\end{document}